\documentclass[aps,prl,reprint,preprintnumbers,showpacs,showkeys,superscriptaddress,nofootinbib,floatfix]{revtex4-1}
\usepackage{amsmath,amssymb,bm,mathrsfs}
\usepackage{times}
\usepackage{epsfig}
\usepackage[colorlinks=true,citecolor=blue,linkcolor=blue]{hyperref}

\usepackage{subfigure}
\usepackage{graphicx}
\usepackage{multirow}

\usepackage{tikz}
\usetikzlibrary{positioning,decorations.pathreplacing,decorations.markings,shapes}
\usetikzlibrary{calc}
\usetikzlibrary{arrows,shapes,backgrounds}
\usetikzlibrary{decorations.pathmorphing}
\usetikzlibrary{decorations.markings}

\makeatletter
\newsavebox\myboxA
\newsavebox\myboxB
\newlength\mylenA
\newcommand*\xoverline[2][0.75]{%
    \sbox{\myboxA}{$\m@th#2$}%
    \setbox\myboxB\null
    \ht\myboxB=\ht\myboxA%
    \dp\myboxB=\dp\myboxA%
    \wd\myboxB=#1\wd\myboxA
    \sbox\myboxB{$\m@th\overline{\copy\myboxB}$}
    \setlength\mylenA{\the\wd\myboxA}
    \addtolength\mylenA{-\the\wd\myboxB}%
    \ifdim\wd\myboxB<\wd\myboxA%
       \rlap{\hskip 0.5\mylenA\usebox\myboxB}{\usebox\myboxA}%
    \else
        \hskip -0.5\mylenA\rlap{\usebox\myboxA}{\hskip 0.5\mylenA\usebox\myboxB}%
    \fi}
\makeatother

\def\pd{\partial}

\def\abs#1{\left| #1 \right|}

\def\a{\alpha}
\def\b{\beta}
\def\d{\delta}
\def\D{\Delta}
\def\g{\gamma}

\def\e{\epsilon}

\def\L{\Lambda}
\def\m{\mu}

\def\scL{\mathcal{L}}

\def\scO{\mathcal{O}}

\def\b4{\xoverline{4}}

\def\b{\beta}

\newcommand{\Dfb}{\mathord{\buildrel{\lower3pt\hbox{$\scriptscriptstyle{\leftrightarrow \tiny{ \ \ \ } }$}}\over {D^{\mu}}}} 

\newcommand{\Dfbd}{\mathord{\buildrel{\lower3pt\hbox{$\scriptscriptstyle\leftrightarrow$}}\over {D}_{\mu}}} 

\def\mt{m_{\tilde{t}}}

\begin{document}
\title{What do precision Higgs measurements buy us?}

\author{Brian Henning}
\email{bhenning@berkeley.edu}
\affiliation{Department of Physics, University of California,
  Berkeley, California 94720, USA}
\affiliation{Theoretical Physics Group, Lawrence Berkeley National
  Laboratory, Berkeley, California 94720, USA}

\author{Xiaochuan Lu}
\email{luxiaochuan123456@berkeley.edu}
\affiliation{Department of Physics, University of California,
  Berkeley, California 94720, USA}
\affiliation{Theoretical Physics Group, Lawrence Berkeley National
  Laboratory, Berkeley, California 94720, USA}

\author{Hitoshi Murayama}
\email{hitoshi@berkeley.edu, hitoshi.murayama@ipmu.jp}
\affiliation{Department of Physics, University of California,
  Berkeley, California 94720, USA}
\affiliation{Theoretical Physics Group, Lawrence Berkeley National
  Laboratory, Berkeley, California 94720, USA}
\affiliation{Kavli Institute for the Physics and Mathematics of the
  Universe (WPI), Todai Institutes for Advanced Study, University of Tokyo,
  Kashiwa 277-8583, Japan}

\begin{abstract}
  We study the sensitivities of future precision Higgs measurements
  and electroweak observables in probing physics beyond the Standard
  Model.  Using effective field theory---appropriate since precision
  measurements are indirect probes of new physics---we examine two
  well-motivated test cases.  One is a tree-level example due to a
  singlet scalar field that enables the first-order electroweak phase
  transition for baryogenesis.  The other is a one-loop example due to
  scalar top in the MSSM.  We find both Higgs and electroweak
  measurements are sensitive probes of these cases.
\end{abstract}
\preprint{UCB-PTH-14/06, IPMU14-0082}
\maketitle

For decades, experimental efforts have chased the Higgs boson like the \emph{holy grail} while, at the same time, theoretical pursuits have tried to make sense of all of its unnatural and mysterious features. Having discovered a ``Higgs
boson''~\cite{Aad:2012tfa,Chatrchyan:2012ufa}, these unnatural and mysterious features immediately become pressing questions. Models of new physics address these questions by making the Higgs more natural if we can avoid a finely-tuned cancellation between the bare parameter and the quadratic divergence
in its mass-sqaured and less mysterious if we can explain why
there is only one scalar in the theory and what dynamics causes it to
condense in the Universe.

Obviously we need to study this new particle as precisely as we can,
which calls for an $e^+ e^-$ collider such as ILC or a circular
machine (TLEP/CEPC).  ILC has been through an intensive internatinonal
study through six-year-long Global Design Effort that released the
Technical Design Report in
2013~\cite{Behnke:2013xla,*Baer:2013cma,*Adolphsen:2013jya,*Adolphsen:2013kya,*Behnke:2013lya}.
Given the technical readiness, we hope to understand the fiscal
readiness in the next few years.  The studies on a very high
intensity circular machine have just started~\cite{Gomez-Ceballos:2013zzn}.

In the past, precision measurements using electrons revealed the next
important energy scale and justified the next big machine.  The
polarized electron-deuteron scattering at SLAC measured the weak
neutral currents precisely~\cite{Prescott:1979dh}, which led to the
justification of Sp$\bar{\rm p}$S and LEP colliders to study $W/Z$
bosons.  The precision measurements at SLC/LEP predicted the mass of
the top quark~\cite{Alexander:1991vi} and the Higgs
boson~\cite{ALEPH:2005ab}, which were verified at the
Tevatron~\cite{Abe:1995hr,Abachi:1995iq} and
LHC~\cite{Aad:2012tfa,Chatrchyan:2012ufa}, respectively. We hope that
precision measurements of the Higgs boson will again point the way to
a definite energy scale.

In this letter, we study what precision Higgs measurements may tell us
for two very different new physics scenarios. One is a singlet scalar
coupled to the Higgs boson, where impacts arise at the tree level. It
can achieve first-order electroweak phase transition which would allow
electroweak baryogenesis. The other is the scalar top in the Minimal
Supersymmetric Standard Model (MSSM), where impacts arise at the
one-loop level. It will help minimize the fine-tuning in the Higgs
mass-squared. In both cases, we find the sensitivities of future
precision Higgs and precision electroweak measurements are similar.

\section{The Standard Model effective field theory}\label{sec:SMEFT}

Precision physics programs offer \emph{indirect} probes of new physics, thereby neccesitating a model-independent framework to analyze potential patterns of deviation from known physics. This framework is most naturally formulated in the language of an effective field theory (EFT) which, for our interests, consists of the Standard Model (SM) supplemented with higher-dimension interactions,
\begin{equation}
\scL_{\text{eff}} = \scL_{\text{SM}} + \sum_i \frac{1}{\L^{d_i-4}} c_i\scO_i. \label{eqn:SM_EFT}
\end{equation}
In the above, \(\L\) is the cutoff scale of the EFT, \(\scO_i\) are dimension \(d_i\) operators that respect the \(SU(3)_c\times SU(2)_L \times U(1)_Y\) gauge invariance of \(\scL_{\text{SM}}\), and \(c_i\) are their Wilson coefficients. In the following, we loosely use the term Wilson coefficient to refer to either \(c_i\) or the operator coefficient, \(c_i/\L^{d_i-4}\). The meaning is clear from context.

Effective field theories are arguably the most appropriate
framework for studying the indirect probes of a precision program.
However, we need to know just how big do we expect the Wilson
coefficients to be in well-motivated models of beyond the Standard
Model (BSM) physics. To shed light on this question, for the models
studied in this letter we first integrate out heavy states and obtain
the Wilson coefficients of the generated higher-dimension operators
and then relate these coefficients to measurable Higgs
observables.

In practice, due to suppression by the high scale \(\L\), the irrelevant operators kept in the EFT are truncated at some dimension. The estimated per mille sensitivity of future precision Higgs programs, together with the present lack of evidence of BSM physics coupled to the SM, justifies keeping only the lowest dimension operators in the effective theory. In the SM effective theory this includes a single dimension-five operator that generates neutrino masses (that we henceforth ignore) and dimension-six operators.

There is a caveat in interpreting Wilson coefficients as the inverse
of heavy particle masses if BSM states couple directly to the Higgs.
The Wilson coefficients in Eq.~\eqref{eqn:SM_EFT} are computed with
mass parameters in the Lagrangian, while the actual mass eigenvalues
receive additional contribution from the Higgs vev and mixings.  This
difference is accounted for by higher-dimension operators which are
dropped in our analysis.  Therefore, the experimental sensitivities on
Wilson coefficients do not translate directly into those on heavy
particle masses. We will quantify this difference in each example.

\renewcommand\arraystretch{1.4}
\begin{table}[tb]
\centering
\begin{tabular}{|rcl|rcl|}\hline
 \(\scO_{GG}\) &\(=\)& \(g_s^2 \abs{H}^2G_{\mu \nu }^aG^{a,\mu \nu }\) & \(\scO_H\)   &\(=\)& \(\frac{1}{2}\big(\pd_{\mu} \abs{H}^2\big)^2\)\\
 \(\scO_{WW}\) &\(=\)& \(g^2  \abs{H}^2 W_{\mu \nu }^aW^{a,\mu \nu } \) &  \(\scO_T\)   &\(=\)& \(\frac{1}{2}\big( H^{\dag} \Dfbd H\big)^2\) \\
 \(\scO_{BB}\) &\(=\)& \(g'^2 \abs{H}^2 B_{\mu \nu }B^{\mu \nu }\) & \(\scO_R\)   &\(=\)& \(\abs{H}^2\abs{D_{\m}H}^2\) \\
 \(\scO_{WB}\) &\(=\)& \(2gg'H^\dag {t^a}H W_{\mu \nu }^a B^{\mu \nu }\) &  \(\scO_D\)   &\(=\)& \(\abs{D^2H}^2\) \\
 \(\scO_W\)   &\(=\)& \(ig\big(H^\dag t^a \Dfb H\big)D^\nu W_{\mu \nu }^a\) &  \(\scO_6\)   &\(=\)& \(\abs{H}^6\) \\
 \(\scO_B\)   &\(=\)& \(ig'Y_H\big(H^\dag \Dfb H\big)\pd^\nu B_{\mu \nu }\)  &  \(\scO_{2G}\) &\(=\)& \(-\frac{1}{2} \big(D^\mu G_{\mu \nu }^a\big)^2\) \\
 \(\scO_{3G}\) &\(=\)& \(\frac{1}{3!}g_sf^{abc}G_\rho ^{a\mu }G_\mu ^{b\nu }G_\nu ^{c\rho }\) &  \(\scO_{2W}\) &\(=\)& \(-\frac{1}{2} \big(D^\mu W_{\mu \nu }^a\big)^2\) \\
 \(\scO_{3W}\) &\(=\)& \(\frac{1}{3!}g \e^{abc}W_\rho ^{a\mu }W_\mu ^{b\nu }W_\nu ^{c\rho }\) & \(\scO_{2B}\) &\(=\)& \(-\frac{1}{2} \big(\pd^{\mu} B_{\mu \nu }\big)^2\) \\
  \hline
\end{tabular}
\caption{\label{tbl:operators} dimension-six bosonic operators for
our analysis.}
\vspace{-10pt}
\end{table}
\renewcommand\arraystretch{0}

We now turn our attention to the dimension-six operators relevant for
our analysis. Since many of the most sensitive probes of Higgs
properties involve only bosons, we restrict our attention to the
purely bosonic dimension-six operators listed in
Table~\ref{tbl:operators}. Some of these operators are redundant
because they can be rewritten by other dimension-six operators using
the SM equations of motion (\emph{e.g.}
\(\scO_{2G}\))~\cite{Buchmuller:1985jz,Grzadkowski:2010es}. We
maintain these so-called redundant operators in our analysis because
(1) their impact on physical observables remains most transparent and
(2) they are directly generated using standard techniques of
integrating out heavy states. While the relationship between some of
these operators and physical observables can be found in the
literature (\emph{e.g.}~\cite{Elias-Miro:2013mua,Elias-Miro:2013eta,Pomarol:2013zra,Alonso:2013hga,Willenbrock:2014bja}),
we provide elsewhere the complete mapping between the operators in
Table~\ref{tbl:operators} and physical observables as well as
techniques for obtaining their Wilson coefficients from UV
models~\cite{HLMfuture:2014}.

Over the past year there has been much progress on understanding the
SM EFT and its relation to Higgs physics. We briefly comment on some
of these developments (see~\cite{Willenbrock:2014bja} for a recent
review). A common theme is the basis of operators in the effective
theory; a complete basis of dimension-six operators contains 59
operators~\cite{Grzadkowski:2010es}.  The choice of this basis is
not unique; however, maintaining a complete basis is crucial for
consistent treatment of renormalization group (RG) evolution within
the EFT~\cite{Jenkins:2013zja}. Several different bases are common in
the
literature~\cite{Grzadkowski:2010es,Giudice:2007fh,Hagiwara:1993ck}
(see~\cite{Willenbrock:2014bja} for comparison), and even these are
often slightly
tweaked~\cite{Elias-Miro:2013mua,Elias-Miro:2013eta}. Our choice of
operators in Table~\ref{tbl:operators} coincides
with~\cite{Elias-Miro:2013eta}, supplemented by the operators
\(\scO_D\) and \(\scO_R\). After specifying a (potentially
overcomplete) basis, the Wilson coefficients can be mapped onto
physical
observables~\cite{Willenbrock:2014bja,Elias-Miro:2013mua,Elias-Miro:2013eta,Pomarol:2013zra,Alonso:2013hga,HLMfuture:2014}. An
overcomplete basis containing redundant operators may also be used,
although the RG evolution requires some
care~\cite{Jenkins:2013zja,Elias-Miro:2013mua,Elias-Miro:2013eta}. Global
fits and constraints on the size of Wilson coefficients in the EFT
have also been
analyzed~\cite{Pomarol:2013zra,Elias-Miro:2013mua,Elias-Miro:2013eta,Chen:2013kfa,*Mebane:2013zga,*Mebane:2013cra}.

\section{A massive singlet}\label{sec:singlet}

We consider a heavy gauge singlet that couples to the SM via a Higgs portal
\begin{eqnarray}
{\cal L} &=& {{\cal L}_{\text{SM}}} + \frac{1}{2}{(\partial_{\mu} S)^2} - \frac{1}{2}m_S^2{S^2} - {A}{\left| H \right|^2}S
\nonumber\\
&& - \frac{1}{2}k{\left| H \right|^2}{S^2} - \frac{1}{{3!}}\mu {S^3} - \frac{1}{{4!}}{\lambda_S^{}}{S^4}. \label{eqn:singlet_L}
\end{eqnarray}
There are several motivations for studying this singlet model. This single additional degree of freedom can successfully achieve a strongly first-order electroweak phase transition (EWPT)~\cite{Grojean:2004xa}. Additionally, singlet sectors of the above form---with particular relations among the couplings---arise in the NMSSM~\cite{Ellis:1988er} and its variants, \emph{e.g.}~\cite{Lu:2013cta,Barbieri:2006bg}. Finally, the effects of Higgs portal operators are captured through the trilinear and quartic interactions \(S \abs{H}^2\) and \(S^2\abs{H}^2\), respectively.

\begin{figure}[tpb]
\centering
\begin{tikzpicture}[>=latex]
\begin{scope}
\node at (0,0) {\includegraphics[width=0.85\linewidth]{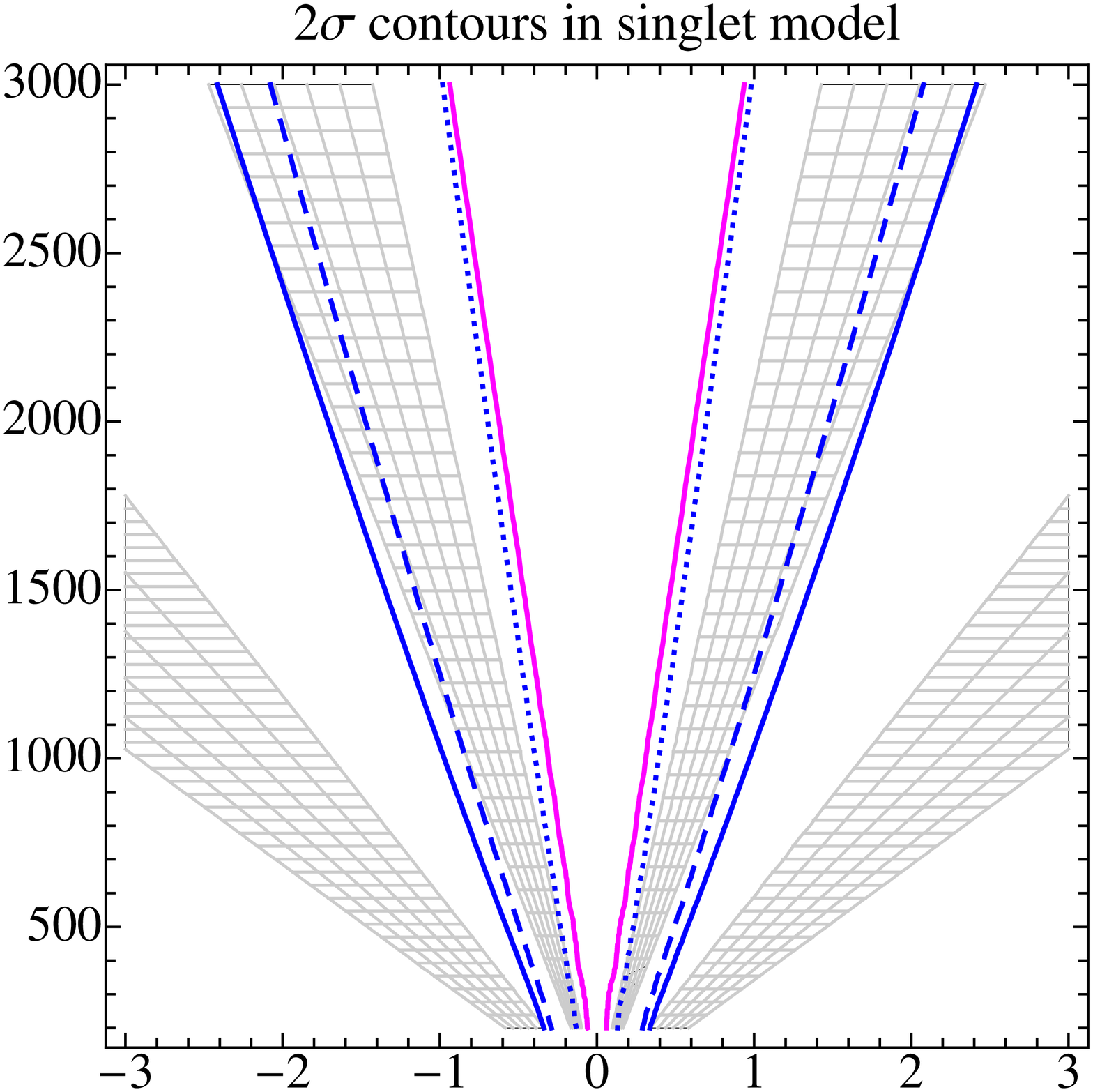}};
\end{scope}

\node at (0.5,-3.7) {\(A/m_S^{}\)};

\node[rotate=90] at (-3.75,0.3) {\(m_{S}^{} \text{ (GeV)}\)};

\node[magenta] at (.5,2) {\footnotesize Oblique};
\draw[->,magenta,thick] (.3,1.8) to [out=-90,in=25] (-.18,1.25);

\node[blue] at (.5,0.7) {\footnotesize\((S,T)\)};
\node[blue] at (.5,0.4) {\scriptsize TeraZ};
\draw[->,dotted,blue,thick] (.5,0.2) to [out=-90,in=155] (.95,-.25);

\node[blue] at (-2.2,1.9) {\footnotesize\((S,T)\)};
\node[blue] at (-2.2,1.6) {\scriptsize GigaZ};
\draw[->,dashed,blue,thick] (-2.25,2.1) to [out=80,in=200] (-1.6,2.7);

\node[blue] at (2.8,.9) {\footnotesize\((S,T)\)};
\node[blue] at (2.8,.6) {\scriptsize Current};
\draw[->,blue,thick] (2.95,1.1) to [out=80,in=-30] (2.5,1.8);

\node[darkgray,rotate=45] at (2.6,-1.2) {\(\boldsymbol{k = 1}\)};

\node[darkgray,rotate=75] at (1.9,1.9) {\(\boldsymbol{k = 4\pi}\)};

\end{tikzpicture}
\vspace{-5pt}
\caption{\label{fig:Singlet} \(2\sigma\) contours of future precision
  measurements on the singlet model in
  Eq.~\eqref{eqn:singlet_L}. Regions below the contours will be
  probed.  The magenta contour is the \(2\sigma\) sensitivity to the
  universal Higgs oblique correction in Eq.~\eqref{eqn:Zh} at ILC
  500up. Blue contours show the \(2\sigma\) RG-induced constraints
  from the \(S\) and \(T\) parameters in
  Eqs.~\eqref{eqn:sing_S}-\eqref{eqn:sing_T} from current measurements
  (solid)~\cite{Baak:2012kk} and future sensitivities at ILC GigaZ
  (dashed)~\cite{Baak:2013fwa} and TLEP TeraZ
  (dotted)~\cite{Mishima:2013}. Regions of a viable first order EW
  phase transition, from Eq.~\eqref{eqn:first_order_PT_bounds}, are
  shown in the gray, hatched regions for \(k=1\) and \(4\pi\).}
\vspace{-10pt}
\end{figure}
For \(m_S^{} \gg m_H\) the singlet can be integrated out; at tree level the resultant low-energy theory contains a finite correction to the Higgs potential as well as the operators \(\scO_H\) and \(\scO_6\):
\begin{equation}
\scL_{\text{eff}} = \scL_{\text{SM}} + \frac{A^2}{2m_S^2} \abs{H}^4 + \frac{A^2}{m_S^4} \scO_H - \bigg(\frac{A^2k}{m_S^4} - \frac{A^3\mu}{m_S^6}\bigg) \scO_6. \label{eqn:singlet_L_eff}
\end{equation}
Upon electroweak symmetry breaking, \(\scO_H\) modifies the wavefunction of the physical Higgs \(h\) and therefore universally modifies all the Higgs couplings,
\begin{equation}
\scL_{\text{eff}} \supset \bigg(1 + \frac{2v^2}{m_S^2}c_H^{}\bigg)\frac{1}{2} (\pd_{\m}h)^2 \Rightarrow \d Z_h = \frac{2v^2}{m_S^2}c_H^{} \label{eqn:Zh},
\end{equation}
where \(c_H^{} = A^2/m_S^2\). This universal {\it Higgs oblique correction} $\delta Z_h$ can be quite sensitive to new physics~\cite{Craig:2013xia,Englert:2013tya,Gori:2013mia} since future lepton colliders, such as the ILC, can probe it at the per mille level~\cite{Dawson:2013bba}. In Fig.~\ref{fig:Singlet}, we show the $2\sigma$ contour of this oblique correction. The contour is obtained by combining the future expected sensitivities of Higgs couplings across all 7 channels in Table 1-20 of~\cite{Dawson:2013bba} for an ILC 500up program, except for the $h\gamma\gamma$ channel where we used the updated value provided by the second column in Table 6 of~\cite{Peskin:2013xra}. As shown, the ILC is quite sensitive to this oblique correction, exploring masses up to several TeV and much of the parameter space of the singlet's couplings to the SM. 

In addition to the oblique correction, \(\scO_H\) will generate measurable contributions to electroweak precision observables (EWPO) under renormalization group evolution. The anomalous dimension matrix \(\g_{ij}\) characterizes the RG mixing amongst dimension-six operators in the SM EFT from a UV scale \(\L\) to the weak scale \(m_W^{}\),
\begin{equation}
c_i(m_W^{}) = c_i(\L) - \frac{1}{16\pi^2} \g_{ij}c_j(\L) \log \frac{\L}{m_W^{}}\ .
\end{equation}
The anomalous dimension matrix has been recently computed~\cite{Elias-Miro:2013mua,Elias-Miro:2013eta,Jenkins:2013zja,Jenkins:2013wua,Alonso:2013hga}. We use the results of~\cite{Elias-Miro:2013eta}.
\footnote{We note that the work~\cite{Elias-Miro:2013eta} calculates \(\g_{ij}\) within a complete operator basis even though they provide only a subset of the full anomalous dimension matrix. Further, upon changing bases, the results of~\cite{Elias-Miro:2013eta} agree with another recent computation of the full anomalous dimension matrix~\cite{Jenkins:2013zja,Jenkins:2013wua,Alonso:2013hga}.} 

Of the EWPO, we find the \(S\) and \(T\) parameters to be the most constraining; in terms of the operators in Table~\ref{tbl:operators} the \(S\) and \(T\) parameters are given by
\begin{align}
S &= \frac{4\sin^2 \theta_W}{\a} \frac{m_W^2}{\L^2} \big[4c_{WB}^{}+ c_W^{}+ c_B^{}\big] (m_W^{}), \label{eqn:S} \\
T &= \frac{1}{\a} \frac{2v^2}{\L^2} c_T^{}(m_W^{}), \label{eqn:T}
\end{align}
where \(v=174 \text{ GeV}\). RG evolution of \(\scO_H\) generates the operators \(\scO_W,\scO_B,\) and \(\scO_T\) with anomalous dimension coefficients~\cite{Elias-Miro:2013eta}
\begin{equation}
\g_{c_H \to c_W} = \g_{c_H \to c_B} = -\frac{1}{3}, \ \g_{c_H \to c_T} = \frac{3}{2}g'^2.
\end{equation}

\begin{figure*}[tpb]
\centering
\begin{tikzpicture}[>=latex]
\begin{scope}
\node at (0,0) {
\subfigure[]{\includegraphics[width=0.393\linewidth]{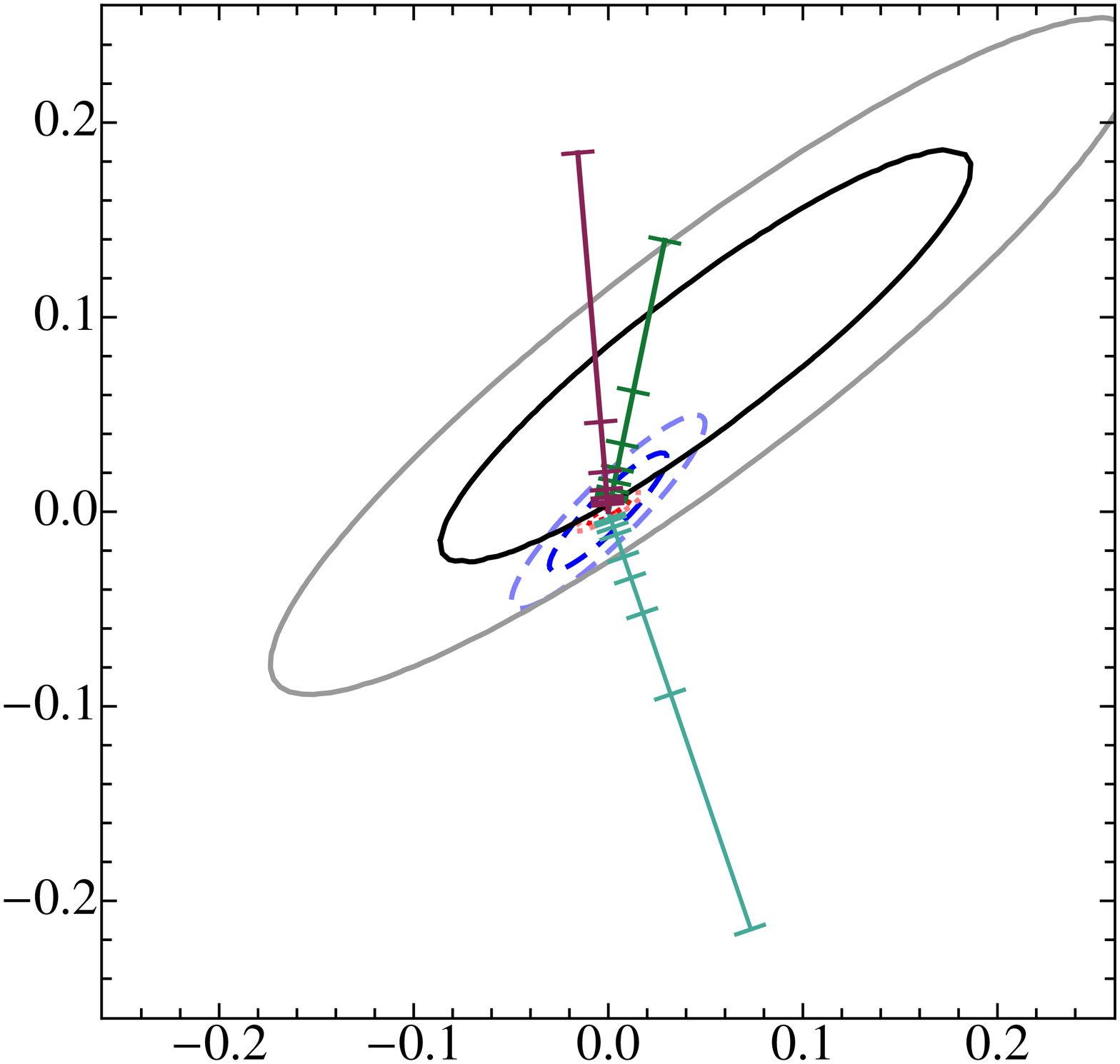}} \(\quad\)
\subfigure[]{\includegraphics[width=0.4\linewidth]{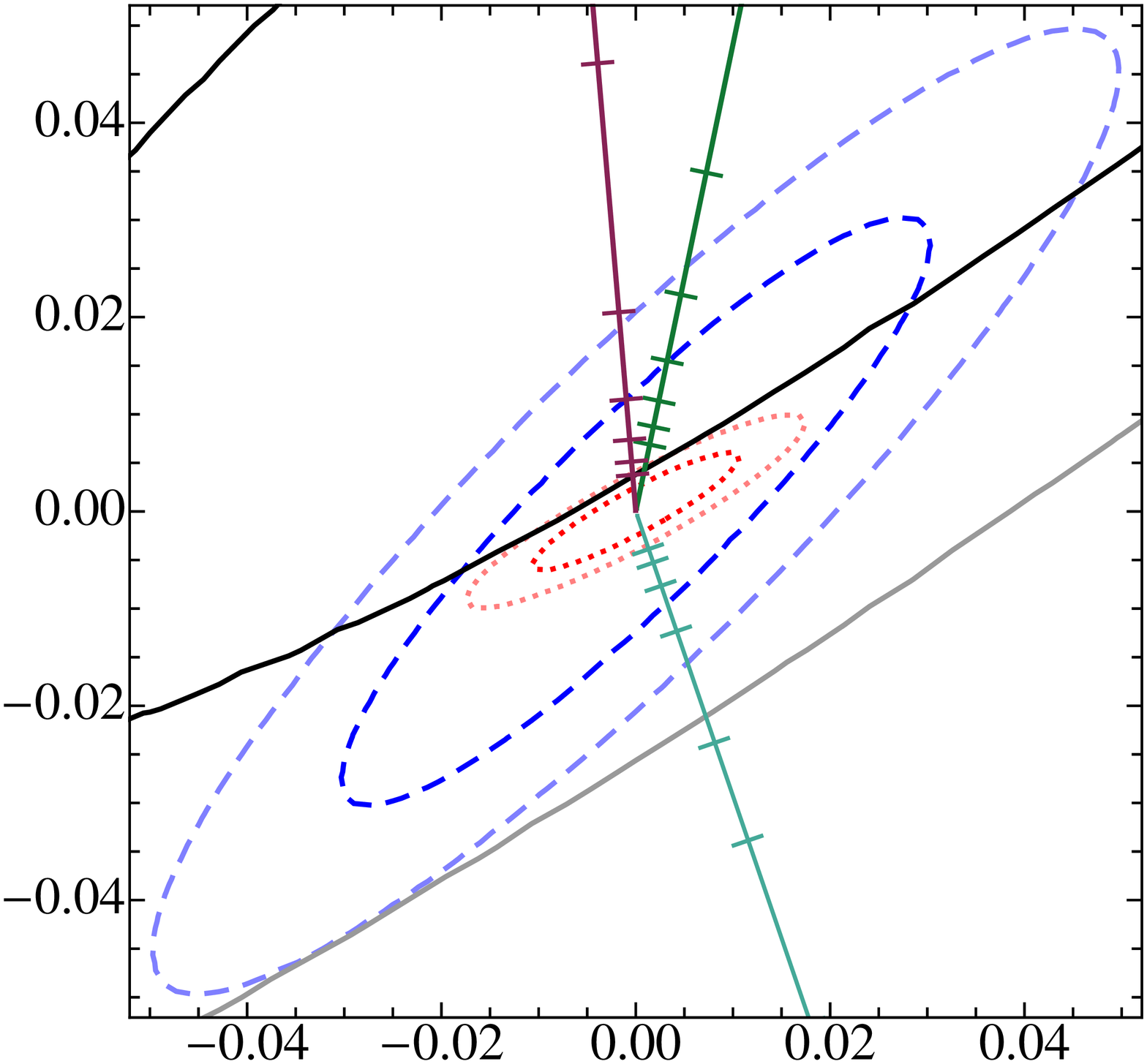}}
};
\end{scope}

\node at (-3.43,-3.2) {\large \(S\)};
\node[rotate=90] at (-7.3,0.5) {\large \(T\)};

\node at (4.23,-3.2) {\large \(S\)};
\node[rotate=90] at (0.2,0.5) {\large \(T\)};

\definecolor{singcol}{RGB}{68,170,153};
\definecolor{Xtcol}{RGB}{17,119,51};
\definecolor{MScol}{RGB}{136,34,85};

\draw[-,thick]             (-6.,-1.3)--(-5.5,-1.3);
\draw[-,thick,dashed,blue] (-6.,-1.7)--(-5.5,-1.7);
\draw[-,thick,dotted,red]  (-6.,-2.1)--(-5.5,-2.1);

\node[anchor=west] at (-5.5,-1.3) {\scriptsize{\bf{Present}}};
\node[anchor=west,blue] at (-5.5,-1.7) {\scriptsize{\bf{GigaZ}}};
\node[anchor=west,red] at (-5.5,-2.1) {\scriptsize{\bf{TeraZ}}};

\node[MScol] at (-4.3,2.2) {\scriptsize{ \bf MSSM}};
\node[MScol] at (-4.3,1.9) {\scriptsize{ \(\boldsymbol{X_t = 0}\)}};

\node[Xtcol] at (-2.4,3.1) {\scriptsize{ \bf MSSM}};
\node[Xtcol] at (-2.4,2.8) {\scriptsize{ \(\boldsymbol{X_t /m_{\tilde{t}} = \sqrt{6}}\)}};

\node[singcol] at (-2,-.7) {\scriptsize{ \bf{Singlet}}};
\node[singcol] at (-2,-1) {\scriptsize{ \(\boldsymbol{A / m_{S}^{} = 1}\)}};

\def \rotS {18.9};
\def \rotX {-11.75};
\def \rotM {4.85};

\node[singcol,rotate=\rotS] at (-2.23,-1.97) {\scriptsize{\bf 200}};
\node[singcol,rotate=\rotS] at (-2.7,-.55) {\scriptsize{\bf 400}};
\node[singcol,rotate=\rotS] at (-2.88,-.08) {\scriptsize{\bf 600}};

\node[MScol,rotate=\rotM] at (-3.95,2.6) {\scriptsize{\bf 200}};
\node[MScol,rotate=\rotM] at (-3.75,.93) {\scriptsize{\bf 400}};

\node[Xtcol,rotate=\rotX] at (-2.75,2.1) {\scriptsize{\bf 400}};
\node[Xtcol,rotate=\rotX] at (-2.95,1.13) {\scriptsize{\bf 600}};


\draw[-,thick]             (1.7,2.6)--(2.2,2.6);
\draw[-,thick,dashed,blue] (1.7,2.2)--(2.2,2.2);
\draw[-,thick,dotted,red]  (1.7,1.8)--(2.2,1.8);

\node[anchor=west] at      (2.2,2.6) {\scriptsize{\bf{Present}}};
\node[anchor=west,blue] at (2.2,2.2) {\scriptsize{\bf{GigaZ}}};
\node[anchor=west,red] at  (2.2,1.8) {\scriptsize{\bf{TeraZ}}};

\node[MScol,rotate=\rotM] at (3.65,3.15) {\scriptsize{\bf 400}};
\node[MScol,rotate=\rotM] at (3.8,1.65) {\scriptsize{\bf 600}};
\node[MScol,rotate=\rotM] at (3.85,1.1) {\scriptsize{\bf 800}};

\node[Xtcol,rotate=\rotX] at (5.,2.45) {\scriptsize{\bf 800}};
\node[Xtcol,rotate=\rotX] at (4.88,1.7) {\scriptsize{\bf 1000}};
\node[Xtcol,rotate=\rotX] at (4.8,1.3) {\scriptsize{\bf 1200}};

\node[singcol,rotate=\rotS] at (5.27,-1.45) {\scriptsize{\bf 800}};
\node[singcol,rotate=\rotS] at (5.1,-.85) {\scriptsize{\bf 1000}};
\node[singcol,rotate=\rotS] at (4.85,-.15) {\scriptsize{\bf 1500}};

\end{tikzpicture}
\vspace{-10pt}
\caption{\label{fig:ellipse} The \(1\sigma\) (darker) and \(2\sigma\)
  (lighter) ellipses of precision EW parameters \(S\) and \(T\). We
  show current fits (solid, black) together with projected
  sensitivities at ILC GigaZ (dashed, blue) and TLEP TeraZ (dotted,
  red).  The lines show the size of \(S\) and \(T\) parameters in our
  singlet model with \(A = m_S^{}\) (teal) and in the MSSM with \(\tan
  \beta = 30\) for \(X_t = \sqrt{6} \mt\) (green) and \(X_t = 0\)
  (purple). The tick marks show specific mass values in each model; 
  \(\mt\) values in \(200 \text{ GeV}\)
  increments starting from \(\mt = 200 \text{ GeV}\) for \(X_t =0\)
  and \(\mt = 400 \text{ GeV}\) for \(X_t = \sqrt{6} \mt\) in the MSSM;
  \(m_S^{}\) values in 200 GeV
  increments between \(200\)-\(1000 \text{ GeV}\) and \(500 \text{
    GeV}\) increments between \(1000\)-\(3000 \text{
    GeV}\) in the singlet model.}
\vspace{-5pt}
\end{figure*}

For the singlet model at hand,
\begin{align}
S &= \frac{1}{6\pi}\Big[\frac{2v^2}{m_S^2} c_H^{}(m_S^{})\Big] \log \frac{m_S^{}}{m_W^{}}\ , \label{eqn:sing_S}\\
T &= -\frac{3}{8\pi \cos^2 \theta_W}\Big[ \frac{2v^2}{m_S^2} c_H^{}(m_S^{})\Big] \log \frac{m_S^{}}{m_W^{}} \ . \label{eqn:sing_T}
\end{align}
It is worth noting that \(S\) and \(T\) are highly correlated---current fits find a correlation coefficient of \(+0.91\)~\cite{Baak:2012kk}---while the RG evolution of \(c_H^{}\) generates \(S\) and \(T\) in the orthogonal direction of this correlation, as depicted in Fig.~\ref{fig:ellipse}. This orthogonality feature enhances the sensitivity of EWPO to oblique Higgs corrections, even when the new physics does not directly couple to the EW sector.

The current best fit of the \(S\) and \(T\) parameters are~\cite{Baak:2012kk}
\begin{equation}
S = 0.05 \pm 0.09, \quad T = 0.08 \pm 0.07 \ .
\end{equation}
This precision is already sensitive to potential next-to-leading order
physics which typically comes with a loop suppresion, as in our
singlet model. Future lepton colliders will significantly increase the
precision measurements of \(S\) and \(T\); a GigaZ program at the ILC
would increase precision to \(\D S = \D T =
0.02\)~\cite{Baer:2013cma,Baak:2013fwa} while a TeraZ program at TLEP
estimates precision of \(\D S =0.007, \ \D T =
0.004\)~\cite{Gomez-Ceballos:2013zzn,Mishima:2013}. Constraints on our
singlet model from current and prospective future lepton collider
measurements of \(S\) and \(T\) are shown in
Fig.~\ref{fig:Singlet}. As seen in the figure, the combination of
increased precision measurements together with the fact that the
singlet generates \(S\) and \(T\) in the anti-correlated direction,
makes these EWPO a particularly sensitive probe of the singlet.  Note
that the apparent lack of improvement by GigaZ is an artifact of
current non-zero central values in $S$ and $T$.

As previously mentioned, this simple singlet model can achieve a strongly first-order EW phase transition. Essentially, this occurs by having a negative quartic Higgs coupling while stabalizing the potential with \(\scO_6\),
\[
V_H \sim a_2\abs{H}^2 - a_4\abs{H}^4 + a_6\abs{H}^6,
\]
for positive coeffiecients \(a_{4,6}\). Within a thermal mass approximation,\footnote{
A full one-loop calculation at finite temperature does not drastically alter the bounds in Eq.~\eqref{eqn:first_order_PT_bounds}; the lower bound remains the same, while the upper bound is numerically raised by about 25\%~\cite{Delaunay:2007wb}. This region is still well probed by future lepton colliders.} a first-order EWPT occurs when~\cite{Grojean:2004xa}
\begin{equation}
\frac{4v^4}{m_H^2} < \frac{2m_s^4}{k A^2} < \frac{12v^4}{m_H^2} \ ,
\label{eqn:first_order_PT_bounds}
\end{equation}
where we have set \(\m = 0\) for simplicity. The lower bound comes from requiring EW symmetry breaking at zero temperature, while the upper bound comes from requiring \(a_4>0\), which guarantees the phase transition is first order.

The region of viability for a strongly first-order EWPT within the singlet model is shown in Fig.~\ref{fig:Singlet}, for nominal values of the coupling \(k\) (note that \(k\) has an upper limit of \(k \lesssim 4\pi\) from perturbativity and lower limit \(k>0\) from stability). Current EWPO already constrain a substantial fraction of the viable parameter space, while future lepton colliders will probe the entire parameter space.

Finally, we comment on the accuracy of the present calculation. Upon
EW symmetry breaking, \(H \to v + h/\sqrt{2}\), the singlet gains an
additional contribution to its mass-squared of order \(k v^2\) and
mixes with \(h\). The light eigenstate of this mixing is the physical
Higgs with mass 125 GeV. As discussed earlier, these effects make the
mass eigenvalue of the heavy scalar differ from the inverse of the
Wilson coefficient in the effective Lagrangian
Eq.~\eqref{eqn:singlet_L_eff}. The difference is of the order of
\begin{equation*}
\frac{kv^2}{m_S^2} \times \text{max}\Big[1,\frac{A^2}{m_S^2}\Big].
\end{equation*}
We note that this difference is very small over most of the region shown in Fig.~\ref{fig:Singlet}. 

\section{Light scalar tops}\label{sec:MSSM}

As a second benchmark scenario, we consider the MSSM with light scalar
tops (stops) and examine the low energy EFT resultant from integrating
out these states. Stops hold a priviledged position in alleviating the
naturalness problem, \emph{e.g.}~\cite{Papucci:2011wy}. This motivates
us to consider a spectrum with light stops while other supersymmetric
partners are decoupled. Since the stops carry all SM gauge quantum
numbers, all of the dimension-six operators in
Table~\ref{tbl:operators} are generated at leading order
(1-loop). Therefore, they also serve as an excellent computational
example to estimate the parametric size of Wilson coefficients of the
operators in Table~\ref{tbl:operators} resultant from heavy scalar
particles with SM quantum numbers. Since the Wilson coefficients are
generated at 1-loop leading order, we discard, as an approximation,
the relatively smaller RG running effects (2-loop) of the Wilson
coefficients.
\begin{figure}[tpb]
\centering
\begin{tikzpicture}[>=latex]
\begin{scope}
\node at (0,0) {\includegraphics[width=0.9\linewidth]{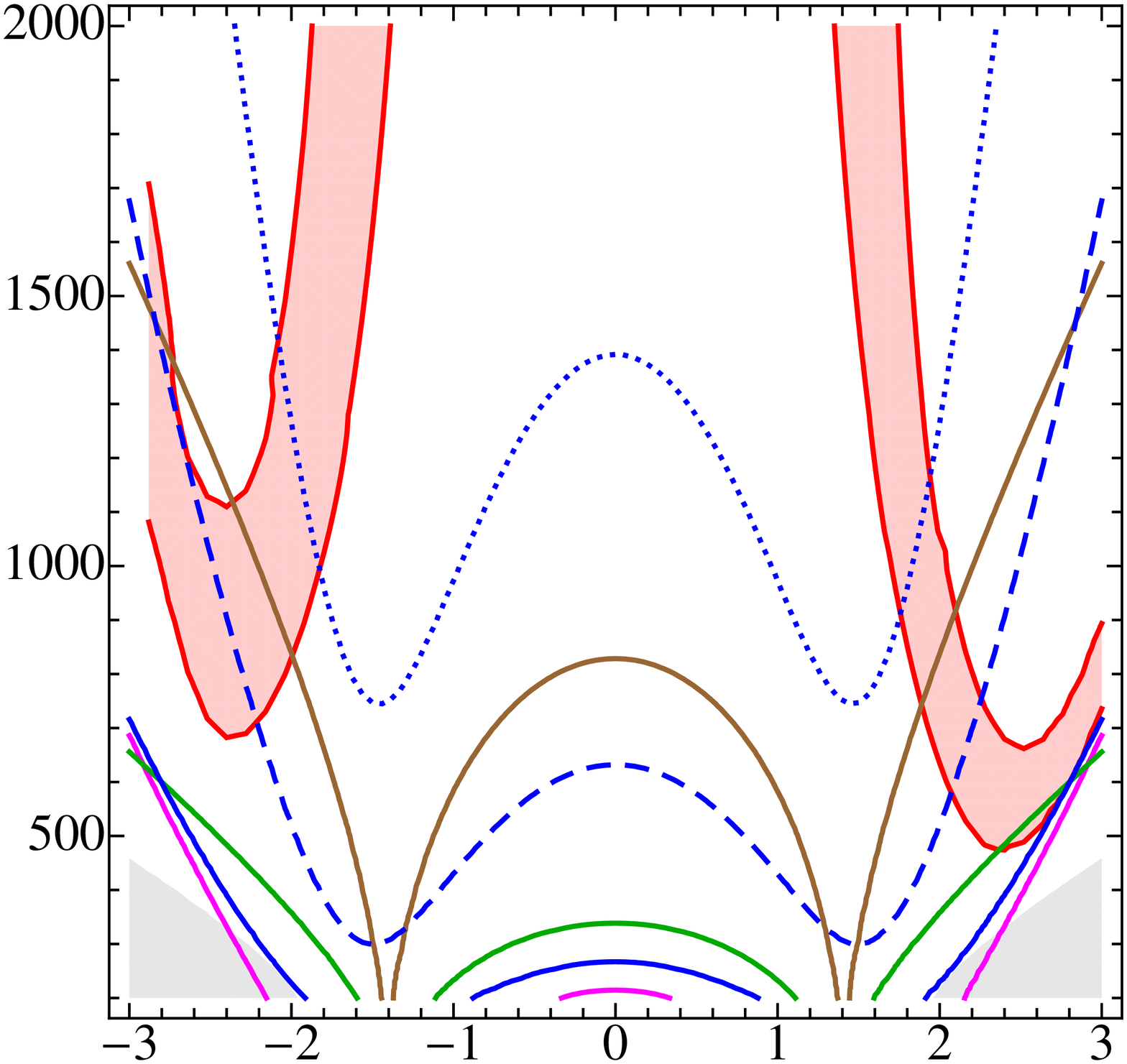}};
\end{scope}

\node at (.5,3.2) {\large{\(\boldsymbol{2\sigma}\) \textbf{contours}}};
\node at (.5,2.75) {\large{\textbf{in MSSM}}};

\node at (.5,2.25) {\small{\(\tan \beta = 30\)}};

\node at (.5,-3.8) {\(X_t/m_{\tilde{t}}\)};
\node[rotate=90] at (-3.85,0.) {\(m_{\tilde{t}} \ \ \text{(GeV)}\)};


\node[blue] at (.5,1.7) {\footnotesize\((S,T)\)};
\node[blue] at (.5,1.4) {\scriptsize TeraZ};
\draw[->,dotted,blue,thick] (.95,1.6) to [out=-15,in=35] (1.1,.85);

\node[blue] at (3.4,3.05) {\footnotesize\((S,T)\)};
\node[blue] at (3.4,2.7) {\scriptsize GigaZ};
\draw[->,dashed,blue,thick] (3.2,2.55) to [out=-80,in=150] (3.6,1.9);

\node[blue] at (3.45,-.75) {\footnotesize\((S,T)\)};
\node[blue] at (3.45,-1.05) {\scriptsize Current};
\draw[->,blue,thick] (3.15,-1.2) to [out=-100,in=190] (3.65,-1.35);

\node[brown!75!black] at (.5,-.6) {\footnotesize \(\boldsymbol{h\to gg}\)};
\draw[->,brown!75!black,thick] (-0.08,-.6) to [out=-170,in=115] (-.3,-1.18);

\node[green!50!black] at (.5,-2.3) {\footnotesize{\(\boldsymbol{h\to \gamma\gamma}\)}};
\draw[->,green!50!black,thick] (1.1,-2.3) to [out=-10,in=75] (1.27,-2.75);

\node[magenta] at (-2.45,-2.95) {\footnotesize Oblique};
\draw[->,magenta,thick] (-2.6,-2.75) to [out=110,in=-120] (-2.5,-2.);

\node[black,rotate=83] at (-1.75,2.85) {\scriptsize{\textbf{127 GeV}}};

\node[black,rotate=83] at (-1.25,2.55) {\scriptsize{\textbf{124 GeV}}};

\end{tikzpicture}
\vspace{-5pt}
\caption{\label{fig:MSSM_tikz} \(2\sigma\) contours of precision Higgs
  and EW observables as a function of \(\mt\) and \(X_t\) in the
  MSSM. The contours show \(2\sigma\) sensitivity of ILC 500up to the
  universal Higgs oblique correction (magenta) and modifications of
  \(h\to gg\) (brown) and \(h \to \gamma\gamma\) (green).  Constraints from
  \(S\) and \(T\) parameters are shown in blue for current
  measurements (solid), ILC GigaZ (dashed), and TLEP TeraZ
  (dotted). The shaded red region shows contours of Higgs mass between
  124-127 GeV in the
  MSSM~\cite{Heinemeyer:1998yj,*Heinemeyer:1998np,*Degrassi:2002fi,*Frank:2006yh}. The
  shaded gray regions are unphysical because one of the stop mass
  eigenvalues becomes negative.}
\vspace{-10pt}
\end{figure}

When we integrate out the multiplet $\phi=(\tilde{Q}_3,{\tilde
  t}_R)^T$, we take degenerate soft masses $m_{{\tilde
    Q}_3}^2=m_{{\tilde t}_R}^2\equiv m_{\tilde t}^2$ for
simplicity. We computed the Wilson coefficients using a covariant
derivative
expansion~\cite{Gaillard:1985uh,Cheyette:1987qz,HLMfuture:2014}
and checked them against standard Feynman diagram
techniques. The resultant Wilson coefficients are listed in
Table~\ref{tbl:MSSM}, where $h_t\equiv m_t/v$ and $X_t = A_t - \mu
\cot \beta$.

As in the previously considered singlet model, these Wilson
coefficients will correct Higgs widths universally through
Eq.~\eqref{eqn:Zh}, as well as contribute to $S$ and $T$ parameters
through Eq.~\eqref{eqn:S}-\eqref{eqn:T}.  In contrast to the singlet
case, the stops contribute to both the oblique correction (via
\(\scO_H\)) and EWPOs (via $\scO_{WB}$, $\scO_W$, $\scO_B$ and
$\scO_T$) at leading order (1-loop). Additionally, vertex corrections
to $h\to gg$ and $h\to \gamma\gamma$ decay widths---arising from
$\scO_{GG}$, $\scO_{WW}$, $\scO_{BB}$, and $\scO_{WB}$---are sensitive
probes since these are loop-level processes within the SM. The
deviations from the SM decay rates are given by
\begin{eqnarray}
 \epsilon_{hgg} &\equiv& \frac{\Gamma _{hgg}}{\Gamma _{hgg}^{\text{SM}}} - 1 = \frac{(4\pi )^2}{\text{Re}A_{hgg}^{\text{SM}}} \frac{16v^2}{\mt^2}c_{GG}^{}, \label{eqngg} \\
 \epsilon_{h\gamma\gamma} &\equiv& \frac{\Gamma _{h\gamma \gamma }}{\Gamma _{h\gamma \gamma}^{\text{SM}}} - 1 
 = \frac{(4\pi )^2}{\text{Re}A_{h\gamma\gamma}^{\text{SM}}}
 \frac{8v^2}{\mt^2} (c_{WW}^{} + c_{BB}^{} -
 c_{WB}^{}), \label{eqngamma} \nonumber \\
\end{eqnarray}
where \(A_{hgg}^{\text{SM}}\) and \(A_{h\g\g}^{\text{SM}}\) are the standard form factors in their respective SM decay rates (see, \emph{e.g.},~\cite{Djouadi:2005gi}).

$2\sigma$ sensitivity contours are shown in
Fig.~\ref{fig:MSSM_tikz}. We stress that here we are focused on the
experimental sensitivities on the scalar top mass, while assuming
improvements on relevant theoretical uncertainties will catch up in
time. Analogous to the case of the singlet model, $m_{\tilde{t}}$ in
the plot differs from the mass eigenvalue by about
$\frac{1}{2}\frac{m_t^2}{m_{\tilde{t}}^2}\times
\max\big(1,\frac{X_t^2}{m_{\tilde{t}}^2}\big)$. As seen in
Fig.~\ref{fig:MSSM_tikz}, future precision Higgs and EW measurements
from the ILC offer comparable sensitivities while a TeraZ program
significantly increases sensitivity. Moreover, the most natural region
of the MSSM---where $X_t\sim\sqrt{6} \mt$ and $m_{\tilde{t}}\sim 1
\text{ TeV}$ (\emph{e.g.}~\cite{Hall:2011aa})---can be well probed by
future precision measurements.

\renewcommand\arraystretch{1.7}
\begin{table*}
\centering
\begin{tikzpicture}[>=latex]
\begin{scope}[xshift=.7cm, yshift = -3.4cm]
\node[anchor=east] at (0,0) {
\scalebox{1.1}{
\begin{tabular}{|rcl|}
  \hline
  \(c_{3G}^{}\) &\(=\)& \(\frac{g_s^2}{(4\pi )^2}\frac{1}{20}\) \\
  \(c_{3W}^{}\) &\(=\)& \(\frac{g^2}{(4\pi)^2}\frac{1}{20}\)  \\
  \(c_{2G}^{}\) &\(=\)& \(\frac{g_s^2}{(4\pi )^2}\frac{1}{20}\) \\
  \(c_{2W}^{}\) &\(=\)& \(\frac{g^2}{(4\pi)^2}\frac{1}{20}\)  \\
  \(c_{2B}^{}\) &\(=\)& \(\frac{g'^2}{(4\pi)^2}\frac{1}{20}\)  \\
  \hline
\end{tabular}
}
};
\end{scope}
\begin{scope}[xshift=-1.cm]
\node[anchor=west] at (0,0) {
\scalebox{1.1}{
\begin{tabular}{|rcl|rcl|}
  \hline
  \(c_{GG}^{}\) &\(=\)& \(\frac{h_t^2}{(4\pi)^2} \frac{1}{12} \left[ \left( 1 + \frac{1}{12}\frac{g'^2c_{2\b} }{h_t^2} \right) - \frac{1}{2}\frac{X_t^2}{\mt^2} \right]\) &
  \(c_{WB}^{}\) &\(=\)& \(-\frac{h_t^2}{(4\pi )^2}\frac{1}{24}\left[ \left( 1 + \frac{1}{2}\frac{g^2c_{2\b} }{h_t^2} \right) - \frac{4}{5}\frac{X_t^2}{\mt^2} \right]\)  \\
  \(c_{WW}^{}\) &\(=\)& \(\frac{h_t^2}{(4\pi)^2} \frac{1}{16} \left[ \left( 1 - \frac{1}{6}\frac{g'^2c_{2\b} }{h_t^2} \right) - \frac{2}{5}\frac{X_t^2}{\mt^2} \right]\) &
  \(c_W^{}\) &\(=\)& \(\frac{h_t^2}{(4\pi)^2}\frac{1}{40} \frac{X_t^2}{\mt^2}\) \\
  \(c_{BB}^{}\) &\(=\)& \(\frac{h_t^2}{(4\pi)^2} \frac{17}{144} \left[ \left( 1 + \frac{31}{102}\frac{g'^2c_{2\b} }{h_t^2} \right) - \frac{38}{85}\frac{X_t^2}{\mt^2} \right]\) &
  \(c_B^{}\) &\(=\)& \(\frac{h_t^2}{(4\pi)^2}\frac{1}{40} \frac{X_t^2}{\mt^2}\) \\[.2cm]
  \hline
\end{tabular}
}
};
\end{scope}

\begin{scope}[xshift=.9cm,yshift = -3.4cm]
\node[anchor=west] at (0,0) {
\scalebox{1.1}{
\begin{tabular}{|rcl|}
  \hline
  \(c_H^{}\) &\(=\)& \(\frac{h_t^4}{(4\pi)^2}\frac{3}{4}\left[ \left( 1 + \frac{1}{3}\frac{g'^2c_{2\b} }{h_t^2} + \frac{1}{12} \frac{g'^4c_{2\b}^2}{h_t^4} \right) - \frac{7}{6}\frac{X_t^2}{\mt^2} \left( 1 + \frac{1}{14}\frac{(g^2 + 2g'^2)c_{2\b} }{h_t^2} \right) + \frac{7}{30}\frac{X_t^4}{\mt^2} \right]\) \\
  \(c_T^{}\) &\(=\)& \(\frac{h_t^4}{(4\pi )^2}\frac{1}{4}\left[ \left( 1 + \frac{1}{2}\frac{g^2c_{2\b} }{h_t^2} \right)^2 - \frac{1}{2}\frac{X_t^2}{\mt^2}\left( 1 + \frac{1}{2}\frac{g^2c_{2\b} }{h_t^2} \right) + \frac{1}{10}\frac{X_t^4}{\mt^4} \right]\) \\
  \(c_R^{}\) &\(=\)& \(\frac{h_t^4}{(4\pi)^2}\frac{1}{2}\left[ \left( 1 + \frac{1}{2}\frac{g^2c_{2\b}}{h_t^2} \right)^2 - \frac{3}{2}\frac{X_t^2}{\mt^2}\left( 1 + \frac{1}{12}\frac{(3g^2 + g'^2)c_{2\b}}{h_t^2} \right) + \frac{3}{10}\frac{X_t^4}{\mt^4} \right]\) \\
  \(c_D^{}\) &\(=\)& \(\frac{h_t^2}{(4\pi)^2}\frac{1}{20} \frac{X_t^2}{\mt^2}\) \\[0.2cm]
  \hline
\end{tabular}
}
};
\end{scope}

\begin{scope}[xshift=-3.5cm,yshift = -6.4cm]
\node[anchor=west] at (0,0) {
\scalebox{1.1}{
\begin{tabular}{|rcl|}
  \hline
  \(c_6^{}\) &\(=\)& \(-\frac{h_t^6}{(4\pi )^2}\frac{1}{2}\left\{ \def\arraystretch{2.2} \begin{array}{l} \left[ 1 + \frac{1}{12}\frac{(3g^2 - g'^2)c_{2\beta }}{h_t^2} \right]^3 + \left[ - \frac{1}{12}\frac{(3g^2 + g'^2)c_{2\beta }}{h_t^2} \right]^3 + \left( 1 + \frac{1}{3}\frac{g'^2c_{2\beta }}{h_t^2} \right)^3 \\  - \frac{X_t^2}{\mt^2}\left[ 2\left( 1 + \frac{1}{12}\frac{(3g^2 - g'^2)c_{2\beta }}{h_t^2} \right) \left( 1 + \frac{1}{8}\frac{(g^2 + g'^2)c_{2\beta }}{h_t^2} \right) + \left(1 + \frac{1}{3}\frac{g'^2c_{2\beta }}{h_t^2} \right)^2 \right] + \frac{X_t^4}{\mt^4}\left[1 + \frac{1}{8}\frac{(g^2 + g'^2)c_{2\beta }}{h_t^2} \right] - \frac{X_t^6}{\mt^6}\frac{1}{10} \\ \end{array} \right\}\) \\
  \hline
\end{tabular}
}
};
\end{scope}

\end{tikzpicture}
\vspace{-15pt}
\caption{\label{tbl:MSSM}  Wilson coefficients \(c_i\) for
  the operators \(\scO_i\) in Table~\ref{tbl:operators} generated
  from integrating out MSSM stops with degenerate soft mass
  \(\mt\). \(g_s,g,\) and \(g'\) denote the gauge couplings of
  \(SU(3),SU(2)_L,\) and \(U(1)_Y\), respectively, \(h_t = m_t/v\),
  and \(\tan \b = \langle
  H_u\rangle / \langle H_d \rangle\) in the MSSM.}
\vspace{-5pt}
\end{table*}

\begin{acknowledgments}
  \noindent
  This work was supported by the U.S. DOE under Contract
  DE-AC03-76SF00098, by the NSF under grants PHY-1002399 and
  PHY-1316783.  HM was also supported by the JSPS grant (C) 23540289,
  and by WPI, MEXT, Japan.
\end{acknowledgments}

\bibliography{./bibliography}
\bibliographystyle{apsrev4-1}

\end{document}